\documentclass{iaus}
\usepackage{graphics}

\title[Sulfur Anomaly] 
{The Sulfur Abundance Anomaly in Planetary Nebulae}

\author[Henry et al.]   
{R.B.C. Henry$^1$,
  J.N. Skinner$^1$, K.B. Kwitter$^2$, \and J.B. Milingo$^3$}

\affiliation{$^1$H.L. Dodge Department of Physics \& Astronomy, 
University of Oklahoma, Norman, OK  73019 USA \break email: 
henry,skinner@nhn.ou.edu\\[\affilskip]
$^2$Department of Astronomy, Williams College, Williamstown, MA 
01267 USA, \break email: kkwitter@williams.edu \break $^3$ Department 
of Physics \& Astronomy, Franklin \& Marshall College, Lancaster, PA 
17604 USA \break email: jmilingo@fandm.edu}

\pubyear{2006}
\volume{234}  
\pagerange{}
\date{?? and in revised form ??}
\jname{Planetary Nebulae in our Galaxy and Beyond}
\editors{M. J. Barlow \& R. H. M\'endez, eds.}
\begin{document}
\firstsection
\maketitle


\firstsection 
\section{What Is The Sulfur Anomaly?}

\begin{figure}[h]
\centering
\scalebox{0.35}{%
\includegraphics{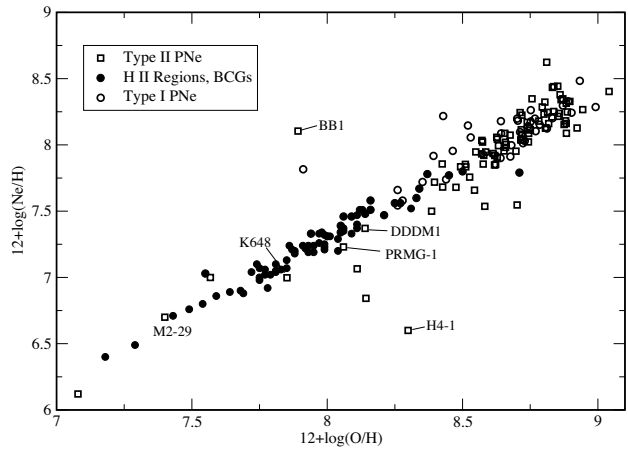}
}
\vskip .05in
\caption{A plot of 12+log(Ne/H) vs. 12+log(O/H).The filled circles 
show data for a merged set of values from the literature for H~II 
regions (M~101; Kennicutt et al. 2003) and blue compact galaxies 
(Izotov \& Thuan 1999). In addition, we show our data for Type~I and 
Type~II PNe as open circles and squares, respectively. Six halo PNe 
included in our data are specifically identified.}
\label{nevso}
\end{figure}

The alpha elements O, Ne, Mg, Si, S, Cl, and Ar are produced together 
in massive stars and their abundances are expected to evolve in 
lockstep. Both H~II regions and planetary nebulae serve as 
interstellar probes of alpha element abundances. We have measured 
abundances for the alpha elements O, Ne, S, Cl, and Ar in over 130 
Galactic Types I and II and halo PNe in a consistent manner from our 
own optical (3600-9600~\AA) spectra (Henry, Kwitter, \& Balick 2004). 
Figure~\ref{nevso} is a plot of Ne vs. O for a combined dataset 
containing H~II regions, blue compact galaxies, and our sample of 
planetary nebulae. All object types appear to be consistent with the 
expected linear relation between Ne and O. Similar plots of Cl and Ar 
vs. O show the same behavior. However, S vs. O is an exception.

\begin{figure}[h]
\centering
\scalebox{0.355}{%
\includegraphics{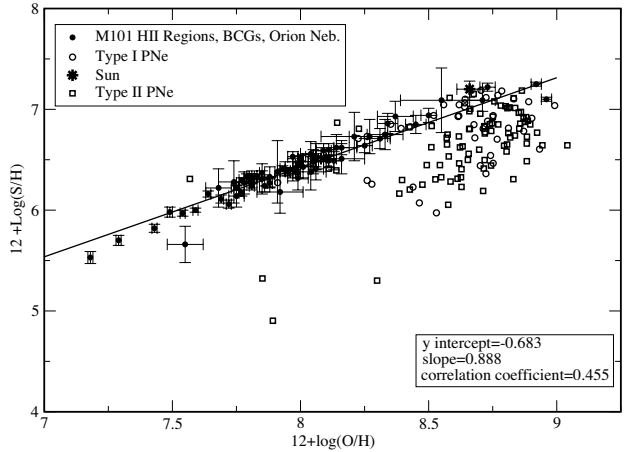}
}
\vskip .05in
\caption{A plot of 12+log(S/H) vs. 12+log(O/H). The symbols are 
the same as in Fig.~\ref{nevso}.}
\label{svso}
\end{figure}

A plot of S vs. O is shown in Fig.~\ref{svso}. The filled circles 
show data for the H~II regions and blue compact galaxies, while the 
open circles and squares represent Type~I and Type~II PNe, 
respectively. The H~II regions and blue compact galaxies and many of 
the Type~I PNe conform to the expected tight linear relation. A 
linear least squares fit to all but the PN data are shown with the 
line; fit parameters are provided in the lower right box. The points 
for the Type~II PNe are highly scattered and fall systematically 
below the linear relation. This unexpected 
behavior between S and O was discovered and discussed in Henry, 
Kwitter, \& Balick (2004) and was dubbed {\underline{\bf the sulfur 
anomaly}}. The same behavior is seen in the S data of 
Kingsburgh \& Barlow (1994) if placed on this plot.

\section{What Causes The Sulfur Anomaly?}

When the vertical distance of a PN from its expected value of S/H 
(the value corresponding to its O/H value in Fig.~\ref{svso}) is plotted as functions of 
its [O~III] temperature, [S~III] temperature, and O$^{+2}$/O$^+$ 
ratio (all measures of nebular excitation), no correlations are 
apparent. The most likely explanation of the sulfur anomaly is that a 
sizeable portion of elemental S is present as S$^{+3}$, which is 
unobservable in the optical. The abundance of S$^{+3}$ is usually accounted for 
through the use of an ionization correction factor (ICF), and we propose 
that abundance methods heretofore have grossly underestimated the 
abundance of this ion. Direct measurements of the S$^{+3}$ abundance 
in PNe can be effected through the observation of [S~IV] 10.5$\mu$m.


Sulfur abundances were determined from ISO data by Pottasch and coworkers 
(Pottasch et al. 2004) using the [S~IV] 10.5$\mu$m line. Seven of 
Pottasch's objects are included in our optical sample, and a comparison of their derived S abundances with ours suggests that in some 
individual cases, the sulfur anomaly is reduced, although the ISO 
data have not solved the problem. The better S/N and overall quality 
of Spitzer data is needed to test our S$^{+3}$ hypothesis in a 
definitive manner. We hope to employ the Spitzer Space Telescope to 
measure the strength of [S~IV] 10.5$\mu$m in 11 Galactic PNe.

We are grateful for support from NSF grant AST 03-07118 and the University of Oklahoma.

\end{document}